\documentclass[aps,prl,reprint,superscriptaddress,nofootinbib]{revtex4-2}
\usepackage{amsmath,amssymb,bm}
\usepackage{graphicx}
\usepackage{siunitx}
\usepackage{microtype}
\usepackage{xcolor}
\usepackage{hyperref}
\usepackage[percent]{overpic}

\begin{document}
%%%%%%%%%%%%%%%%%%%%%%%%%%%%%%%%%%%%%%%%%%%%%%%%%%%%%%%%%%%%%%%%%%%%%%%%%
\title{Thermal scaling laws for open-water swimming}
%%%%%%%%%%%%%%%%%%%%%%%%%%%%%%%%%%%%%%%%%%%%%%%%%%%%%%%%%%%%%%%%%%%%%%%%%
\author{Henry van den Bedem}
%\affiliation{Expedition Medicines, Cambridge, Massachusetts 02141, USA}
\affiliation{Department of Bioengineering and Therapeutic Sciences,
University of California, San Francisco, San Francisco, California 94143, USA}

\author{Ellen Kuhl}
\affiliation{Department of Mechanical Engineering
and Wu Tsai Human Performance Alliance, Stanford University, Stanford, California 94305, USA}
%%%%%%%%%%%%%%%%%%%%%%%%%%%%%%%%%%%%%%%%%%%%%%%%%%%%%%%%%%%%%%%%%%%%%%%%%
\date{\today}
%%%%%%%%%%%%%%%%%%%%%%%%%%%%%%%%%%%%%%%%%%%%%%%%%%%%%%%%%%%%%%%%%%%%%%%%%
\begin{abstract}
Open-water swimming defines a thermal phase-boundary problem in which metabolic heat production competes with environmental heat loss. We derive a scaling law that predicts the critical water temperature and shows how body size, swim pace, and insulation shift this boundary. Longitudinal warm- and cold-water data reveal transient dynamics that exceed single-compartment predictions, but emerge naturally from core--peripheral physiology. Together, our results suggest that thermal safety depends on swimmer-specific characteristics and swimming conditions, not on water temperature alone.
\vspace{0.8em}

\begin{center}
\normalfont\itshape
Dedicated to Cate, Doro, Janine, Marcella, Polly, and all
endurance athletes who have experienced hypothermia.
\end{center}

\end{abstract}
%%%%%%%%%%%%%%%%%%%%%%%%%%%%%%%%%%%%%%%%%%%%%%%%%%%%%%%%%%%%%%%%%%%%%%%%%
\maketitle
%%%%%%%%%%%%%%%%%%%%%%%%%%%%%%%%%%%%%%%%%%%%%%%%%%%%%%%%%%%%%%%%%%%%%%%%%
Emperor penguins ({\it{Aptenodytes forsteri}}) forage in \SI{-1.8}{\degreeCelsius} Antarctic water while maintaining a core temperature near \SI{38}{\degreeCelsius} 
\cite{Williams2015}. Humans, by contrast, can develop hypothermia in water nearly  \SI{20}{\degreeCelsius} warmer \cite{TiptonBradford2014}. This striking disparity exposes a fundamental heat-transfer problem: water temperature alone does not determine thermal state \cite{Nadel1984}. Metabolic heat production competes with environmental heat loss, while body size, insulation, and physiology regulate the exchange between 
body and water \cite{TiptonBradford2014}. Classical calorimetric experiments established the competition between metabolic heat production and environmental heat loss in immersed humans \cite{Cannon1960}, while subsequent swimming experiments connected internal temperature to water temperature, exercise intensity, and body composition \cite{Nadel1974}.

Open-water swimming turns this balance into a paradox: the same water temperature can cool one swimmer while heating another \cite{Markey2026}. Long exposures can drive swimmers toward either hypothermia or hyperthermia \cite{TiptonBradford2014}. Cold-water immersion can impair neuromuscular and cognitive function before deep hypothermia develops \cite{Tipton2017}. Warm water, by contrast, reduces the temperature gradient for heat rejection and can produce substantial thermal strain \cite{Chalmers2021}. Experiments identify metabolic heat production and swimmer morphology as major determinants of individual cooling rates \cite{Saycell2019}. Yet, endpoints, cooling rates, and thermal thresholds alone cannot resolve the \textit{transient trajectory} that connects them.

Competition rules nevertheless characterize thermal exposure primarily through water temperature. World Aquatics defines a competition range from 16 to \SI{31}\degreeCelsius  \cite{WorldAquatics2026}, yet swimmers exposed to similar water temperatures exhibit markedly different thermal responses \cite{Markey2026}. Classical heat-transfer theory relates thermal evolution to heat production, storage, and exchange with the environment \cite{Fourier1822}, while scaling laws characterize heat transport across boundaries \cite{Aoki2001}. Recent experiments identify body fat, metabolic heat production, and surface-area-to-mass ratio as determinants of swimmer-specific thermal 
limits \cite{Skutnik2026}. These observations motivate a shift from universal temperature thresholds toward swimmer-specific thermal boundaries.

Here we formulate this problem through thermal scaling laws and phase 
boundaries. Our central result is the {\it{critical-water-temperature 
scaling law}},
\[
\boxed{T_{\rm w}^{\rm crit} = T_{\rm c}^0 -\frac{Q}{H}}
\label{eq:critical_intro}
\]
which separates 
net cooling for $T_{\rm w} < T_{\rm w}^{\rm crit}$ from 
net heating for $T_{\rm w} > T_{\rm w}^{\rm crit}$,
where $T_{\rm w}^{\rm crit}$ is the critical water temperature,
$T_{\rm c}^0$ is the initial core body temperature,
$Q$ is the metabolic heat production, and
$H$ is the effective thermal conductance.
We derive this law from the fundamental heat balance and extend earlier heat-balance and compartment models through three levels of physics, physiology, and personalization that connect swimmer characteristics to thermal boundaries and core--peripheral timescales. Crucially, we calibrate the models against longitudinal core-temperature trajectories from warm- and cold-water swims rather than initial and final temperatures alone.

%%%%%%%%%%%%%%%%%%%%%%%%%%%%%%%%%%%%%%%%%%%%%%%%%%%%%%%%%%%%%%%%%%%%%%%%%
\begin{figure*}[t]
\centering
\includegraphics[width=\textwidth]{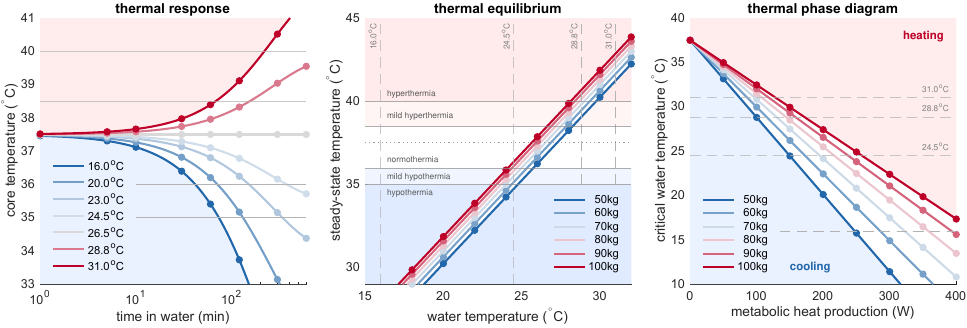}
\caption{\textbf{Thermal scaling laws for open-water swimming.}
Model I predicts core-temperature trajectories from the competition between metabolic heat production and heat loss to water. The same swimmer cools in cold water, approaches a thermal equilibrium near a critical water temperature, and heats in warm water (left).
Steady-state core temperature increases approximately linearly with water temperature and shifts with body mass. Horizontal bands mark hypothermia ($T^{\rm c}<\SI{35}{\degreeCelsius}$), mild hypothermia ($\SI{35}{\degreeCelsius}\le T^{\rm c}<\SI{36}{\degreeCelsius}$), normothermia ($\SI{36}{\degreeCelsius}\le T^{\rm c}<\SI{38.5}{\degreeCelsius}$), mild hyperthermia ($\SI{38.5}{\degreeCelsius}<T^{\rm c}<\SI{40}{\degreeCelsius}$), and hyperthermia ($T^{\rm c}\ge\SI{40}{\degreeCelsius}$) (middle).
The critical water temperature separates cooling from heating and decreases with metabolic heat production. Body mass shifts the boundary since heat capacity and heat-transfer area scale differently with body size (right). 
Solid lines delineate the thermal regions;
dashed lines indicate World Aquatics minimum and maximum competition temperatures at 16.0 and \SI{31.0}{\degreeCelsius} and
Ironman wetsuit-legal and wetsuit-prohibited temperature limits at 24.5 and \SI{28.8}
{\degreeCelsius}.}
\label{fig01}
\end{figure*}
%%%%%%%%%%%%%%%%%%%%%%%%%%%%%%%%%%%%%%%%%%%%%%%%%%%%%%%%%%%%%%%%%%%%%%%%%

\vspace*{0.2cm}
%%%%%%%%%%%%%%%%%%%%%%%%%%%%%%%%%%%%%%%%%%%%%%%%%%%%%%%%%%%%%%%%%%%%%%%%%
\paragraph*{Model I: Physics predicts thermal scaling law.}
%%%%%%%%%%%%%%%%%%%%%%%%%%%%%%%%%%%%%%%%%%%%%%%%%%%%%%%%%%%%%%%%%%%%%%%%%
We first represent the swimmer as a single well-mixed thermal compartment 
with core temperature $T_{\rm c}$, mass $m$, and effective specific heat $c$. 
The balance of energy gives
\begin{equation}
m c\,\dot T_{\rm c} = Q-H\,(T_{\rm c}-T_{\rm w}),
\label{eq:model1}
\end{equation}
where $Q$ is the metabolic heat production, $T_{\rm w}$ is the water 
temperature, $H=h_{\rm eff}A$ is the whole-body thermal conductance, and
$h_{\rm eff}$ is the effective heat-transfer coefficient. 
The surface area $A$ follows the Du Bois scaling relation,
$A=0.007184\,m^{0.425}h^{0.725}$,
with mass $m$ in kilograms and height $h$ in centimeters.
For comparisons between swimmers of different body sizes at the same mass-specific metabolic intensity $q$, we assume $Q=mq$, whereas heat exchange scales with body surface area through $H=h_{\rm eff}A$.
Integrating the energy balance %~(\ref{eq:model1}) 
in time yields
\begin{equation}
T_{\rm c}(t)=T_\infty+(T_{\rm c}^{0}-T_\infty)e^{-t/\tau}
\quad \mbox{with} \quad
\tau=\frac{mc}{H},
\label{eq:solution1}
\end{equation}
where 
$T_\infty$ defines the {\it{steady state temperature}} and
$\tau$ defines the {\it thermal time scale}. 
%{characteristic thermal response time}. 
It represents the ratio between the swimmer's thermal capacitance $mc$ 
and the thermal conductance $H$ to the surrounding water: large 
thermal mass slows the response, strong heat transfer accelerates it.
At {\it{steady state}}, metabolic heat production exactly balances heat transfer 
from swimmer to water, $Q=H \, (T_\infty-T_{\rm w})$, and
\begin{equation}
\boxed{
T_\infty-T_{\rm w}=\frac{Q}{H}
}
\label{eq:steady}
\end{equation}
defines the {\it{metabolic temperature offset}} required to dissipate the heat 
generated during swimming. Notably, the steady-state offset depends 
only on the competition between metabolic heat production $Q$ and environmental 
heat transfer $H$; thermal capacitance $mc$ controls how fast the swimmer approaches 
this state, but not the state itself.
%%%%%%%%%%%%%%%%%%%%%%%%%%%%%%%%%%%%%%%%%%%%%%%%%%%%%%%%%%%%%%%%%%%%%%%%%
\begin{figure*}[t]
\centering
\includegraphics[width=\textwidth]{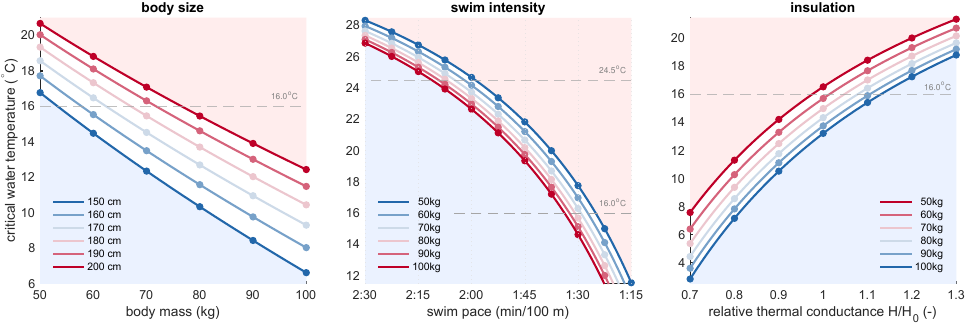}
\caption{\textbf{Personalized critical water temperature.}
Larger body mass lowers the critical water temperature through the ratio of 
heat-producing mass to heat-losing surface area (left).
Faster swim pace increases metabolic heat production and nonlinearly 
lowers the critical water temperature (middle).
Reduced thermal conductance, for example provided by wetsuits, reduces 
environmental heat loss and lowers the critical water temperature (right).
Blue and red regions denote net cooling and heating.
Dashed lines indicate the World Aquatics minimum and maximum competition 
temperatures at 16.0 and \SI{31.0}{\degreeCelsius} and the Ironman 
wetsuit-legal and wetsuit-prohibited temperature limits at 24.5 and 
\SI{28.8}{\degreeCelsius}.}
\label{fig02}
\end{figure*}
%%%%%%%%%%%%%%%%%%%%%%%%%%%%%%%%%%%%%%%%%%%%%%%%%%%%%%%%%%%%%%%%%%%%%%%%%
The steady-state temperature determines the direction of thermal drift:
the swimmer cools for $T_\infty<T_{\rm c}^{0}$, remains in thermal balance 
for $T_\infty=T_{\rm c}^{0}$, and heats for $T_\infty>T_{\rm c}^{0}$. 
Setting $T_\infty=T_{\rm c}^{0}$ defines the 
{\it{critical-water-temperature scaling law}},
\begin{equation}
\boxed{
T_{\rm w}^{\rm crit}
=
T_{\rm c}^{0}-\frac{Q}{H}
}
\label{eq:critical}
\end{equation}
which separates the cooling and heating regimes.
% \noindent 
Figure~\ref{fig01} summarizes three direct consequences of Model~I: 
The left panel evaluates the {\it{transient solution}},
\begin{equation}
  T_{\rm c}(t)
= T_{\rm w}
+ \frac{Q}{H}
+ \left(T_{\rm c}^{0}-T_{\rm w}
- \frac{Q}{H}\right)\exp\!\left(-\frac{H}{mc}t\right),
\label{eq:fig1left}
\end{equation}
where the water temperature $T_{\rm w}$ controls the direction and magnitude of thermal drift, while $\tau=mc/H$ sets its time scale. The middle panel evaluates the steady-state temperature,
$T_\infty(m,T_{\rm w})
=T_{\rm w}
+Q/(h_{\rm eff}\,0.007184\,m^{0.425}h^{0.725})$,
the right panel evaluates the critical water temperature,
$T_{\rm w}^{\rm crit}(m,Q)
=T_{\rm c}^{0}
-Q/(h_{\rm eff}\,0.007184\,m^{0.425}h^{0.725})$.
Increased metabolic heat shifts the cooling--heating boundary toward colder water, while body size shifts it through mass-to-surface-area scaling. 
Notably, fixed race thresholds at 
16.0, 24.5, 28.8, \SI{31.0}{\degreeCelsius}
cut across, rather than follow, these {\it{personalized thermal boundaries}}.
Here, personalization is for a reference swimmer with $m=70\,\mathrm{kg}$, $h=175\,\mathrm{cm}$, and $A=1.85\,\mathrm{m^2}$. We use $H_0=15\,\mathrm{W/K}$, corresponding to $h_{\rm eff,0}=8.11\,\mathrm{W/(m^2K)}$. This effective conductance accounts for the combined influence of tissue insulation, peripheral circulation, and heat exchange with the surrounding water, rather than water-side convection alone. Its magnitude depends on swimming conditions and thermal insulation and is therefore a reference parameter rather than a universal physiological constant.

\vspace*{0.2cm}
%%%%%%%%%%%%%%%%%%%%%%%%%%%%%%%%%%%%%%%%%%%%%%%%%%%%%%%%%%%%%%%%%%%%%%%%%
\paragraph*{Model II: Physiology predicts core--peripheral dynamics.}
%%%%%%%%%%%%%%%%%%%%%%%%%%%%%%%%%%%%%%%%%%%%%%%%%%%%%%%%%%%%%%%%%%%%%%%%%
Model I predicts a monotonic exponential response. Field data reveal 
richer dynamics, including initial plateaus, transient heating, and 
delayed cooling. We therefore partition the swimmer into core and peripheral thermal compartments,
\begin{equation}
\begin{array}{llll}
C_{\rm c} & \dot T_{\rm c} &= \, Q \, - & G\,(T_{\rm c}-T_{\rm p}), \\
C_{\rm p} & \dot T_{\rm p} &=           & G\,(T_{\rm c}-T_{\rm p})-H(T_{\rm p}-T_{\rm w}),
\end{array}
\label{eq:model2}
\end{equation}
where
$C_{\rm c}=\nu_{\rm c} \, mc$ and $C_{\rm p}=(1-\nu_{\rm c}) \, mc$
denote the core and peripheral thermal capacitances,
$\nu_{\rm c}$ and $(1-\nu_{\rm c})$ are the core and peripheral fraction of the total thermal capacitance,
$T_{\rm c}$ and $T_{\rm p}$ are the core and peripheral temperatures, and 
$G$ is the effective core--periphery conductance.
Unlike model I, which gives
$T_{\rm w}^{\rm crit}=T_{\rm c}^{0}-Q/H$,
model II predicts
$T_{\rm c,\infty}=T_{\rm w}+Q/H+Q/G$
and therefore
$T_{\rm w}^{\rm crit}=T_{\rm c}^{0}-Q/H-Q/G$.
Equation (\ref{eq:critical}) applies specifically to model I.
Following classical two-node thermoregulation models \cite{Gagge1971}, this formulation preserves energy balance while introducing a physiological delay between environmental heat exchange and core response. The calibrated $G$ and $\nu_{\rm c}$ represent effective physiological parameters rather than uniquely identified tissue properties; the Appendix reports their calibrated distributions and parameter constraints and compares our reduced-order models with immersion experiments and detailed thermoregulation models \cite{Castellani2023}.

\vspace*{0.2cm}
%%%%%%%%%%%%%%%%%%%%%%%%%%%%%%%%%%%%%%%%%%%%%%%%%%%%%%%%%%%%%%%%%%%%%%%%%
\paragraph*{Model III: Personalization predicts thermal boundaries.}
%%%%%%%%%%%%%%%%%%%%%%%%%%%%%%%%%%%%%%%%%%%%%%%%%%%%%%%%%%%%%%%%%%%%%%%%%
Equation~(\ref{eq:critical}) suggests three independent routes to 
personalization: body size, swim intensity, and insulation. 
Figure~\ref{fig02} varies these mechanisms--one at a time--and highlights their effects on the critical water temperature.
All three panels use the reference swimmer and baseline thermal parameters specified in the Appendix; the curves isolate the effects of body size, swim pace, and insulation on $T_{\rm w}^{\rm crit}$.
The left panel varies mass $m$ and height $h$ as
$T_{\rm w}^{\rm crit}(m,h)
=T_{\rm c}^{0}
-mq
/(h_{\rm eff}\,0.007184\,m^{0.425}h^{0.725})$.
At fixed mass-specific metabolic heat production $q$, 
increasing body size lowers the critical water temperature 
because total heat production scales with mass, $Q=mq$, 
whereas heat loss scales with surface area, $H=h_{\rm eff}A$. 
This mass-to-surface-area dependence is consistent with established heat-transfer models of cold-water immersion \cite{Tarlochan2005}. 
The middle panel varies swim intensity through swim pace $p$, 
the inverse swim velocity $v$ in minutes per 100~m,
and body mass $m$ as
$T_{\rm w}^{\rm crit}(p,m)
=T_{\rm c}^{0}
-\frac{m\,q_{\rm ref}}{h_{\rm eff}A(m,h)}
\left(\frac{100}{60\,p\,v_{\rm ref}}\right)^n$,
where we describe the increase in mass-specific
metabolic heat production through
$q=q_{\rm ref}(v/v_{\rm ref})^n
=q_{\rm ref}(100/(60\,p\,v_{\rm ref}))^n$. 
Faster swimming lowers the pace $p$,
increases metabolic heat production $mq$ nonlinearly, and 
lowers the critical water temperature.
The right panel varies the relative environmental conductance 
$\eta=H/H_0$ and mass $m$ as
$T_{\rm w}^{\rm crit}(\eta,m)
=T_{\rm c}^{0}
-Q/(\eta H_0)$.
Greater insulation reduces the relative thermal heat conductance 
$\eta<1$, reduces heat transfer to 
the surrounding water, and lowers the critical water temperature.
%%%%%%%%%%%%%%%%%%%%%%%%%%%%%%%%%%%%%%%%%%%%%%%%%%%%%%%%%%%%%%%%%%%%%%%%%
\begin{figure*}[t]
\centering
\includegraphics[width=\textwidth]{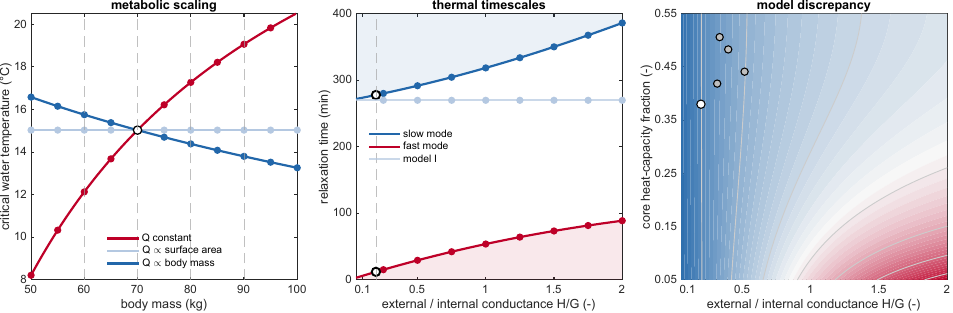}
\caption{\textbf{Metabolic scaling, thermal timescales, and model discrepancy.}
The critical water temperature depends on body mass and the assumed scaling of metabolic heat production: constant total heat production, constant heat production per unit surface area, or constant heat production per unit body mass (left). 
Model I has a single relaxation time, whereas
Model II introduces fast and slow thermal relaxation modes that depend on the ratio of external to internal conductance $(H/G)$ (middle). 
The discrepancy between Models I and II varies with $(H/G)$ and the core heat-capacity fraction $\nu_{\rm c}$ and highlights the regimes in which core--peripheral coupling becomes important (right).
The white circle marks the reference swimmer, and the gray circles indicate the cold-water swimmers analyzed in Figure \ref{fig05}.}
\label{fig03}
\end{figure*}
%%%%%%%%%%%%%%%%%%%%%%%%%%%%%%%%%%%%%%%%%%%%%%%%%%%%%%%%%%%%%%%%%%%%%%%%%
Taken together, larger body size, faster swim pace, and greater 
insulation all lower the critical water temperature: larger swimmers 
retain metabolic heat more effectively, faster swimmers generate more 
metabolic heat, and greater insulation reduces environmental heat loss.
These dependencies challenge the notion that a single water-temperature threshold can define thermal safety for {\it{all}} swimmers. 
Fixed World Aquatics and Ironman thresholds provide practical {\it{population-level guidelines}}, 
but thermal safety ultimately depends on 
{\it{swimmer-specific characteristics}} and swimming conditions.
\vspace*{0.2cm}
%%%%%%%%%%%%%%%%%%%%%%%%%%%%%%%%%%%%%%%%%%%%%%%%%%%%%%%%%%%%%%%%%%%%%%%%%
\paragraph*{Physical scaling and model complexity.} 
%%%%%%%%%%%%%%%%%%%%%%%%%%%%%%%%%%%%%%%%%%%%%%%%%%%%%%%%%%%%%%%%%%%%%%%%%
Figure \ref{fig03} extends the analysis from thermal boundaries to the dynamics of core--peripheral coupling. The left panel compares three metabolic scaling assumptions: constant heat production, constant heat production per unit surface area, and constant heat production per unit mass. The middle panel identifies the fast and slow thermal relaxation times as functions of the conductance ratio $(H/G)$, while the right panel quantifies the discrepancy between Models I and II as functions of $(H/G)$ and the core thermal fraction $\nu_{\rm c}$. Together, these results identify when a single thermal timescale adequately describes the response and when core--peripheral coupling introduces additional dynamics.
\vspace*{0.2cm}
%%%%%%%%%%%%%%%%%%%%%%%%%%%%%%%%%%%%%%%%%%%%%%%%%%%%%%%%%%%%%%%%%%%%%%%%%
\begin{figure*}[t]
\centering
\includegraphics[width=\textwidth]{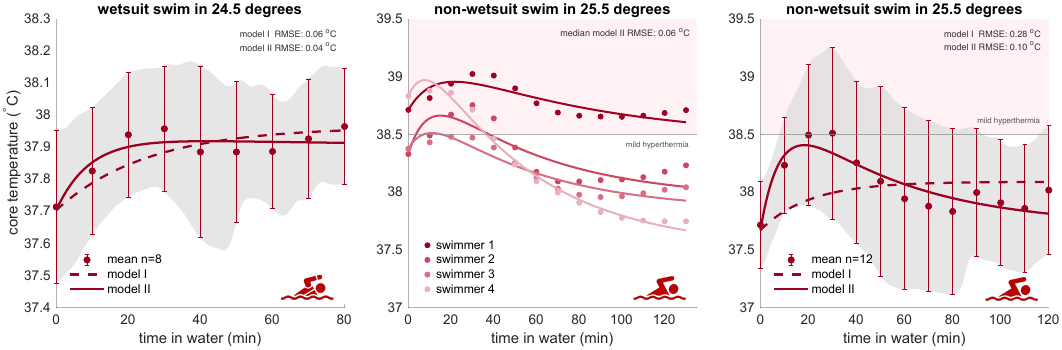}
%\begin{overpic}[width=\textwidth]{fig3.pdf}
%\put(26.4,4.5){\includegraphics[width=0.05\textwidth]{wetsuit_red_trans.png}}
%\put(59.7,4.5){\includegraphics[width=0.05\textwidth]{swimmer_red_trans.png}}
%\put(93.0,4.4){\includegraphics[width=0.05\textwidth]{swimmer_red_trans.png}}
%\end{overpic}
\caption{\textbf{Core temperature dynamics during warm-water swim.}
Dots and shaded region show the mean and standard deviation of n=8 
wetsuit swims in \SI{24.5}{\degreeCelsius} \cite{Morton2025}; 
dashed and solid curves show models I and II 
with {\sf{RMSE}} of 0.06 and \SI{0.04}{\degreeCelsius} (left).
Dots show n=4 representative 
non-wetsuit swims in \SI{25.5}{\degreeCelsius} \cite{Markey2025};
solid curves show that model II 
captures personalized transient heating and subsequent relaxation 
with a median {\sf{RMSE}} of \SI{0.06}{\degreeCelsius} (middle).
Dots and shaded region show the mean and standard deviation of n=12 
non-wetsuit swims in \SI{25.5}{\degreeCelsius} \cite{Markey2025}; 
dashed and solid curves show that model I 
fails to capture the early temperature peak 
while model II reproduces this effect
with {\sf{RMSE}} of 0.28 and \SI{0.10}{\degreeCelsius} (right).
Horizontal lines mark mild hyperthermia at \SI{38.5}{\degreeCelsius}.}
\label{fig04}
\end{figure*}
%%%%%%%%%%%%%%%%%%%%%%%%%%%%%%%%%%%%%%%%%%%%%%%%%%%%%%%%%%%%%%%%%%%%%%%%%
%%%%%%%%%%%%%%%%%%%%%%%%%%%%%%%%%%%%%%%%%%%%%%%%%%%%%%%%%%%%%%%%%%%%%%%%%
\paragraph*{Longitudinal thermal trajectories.}
%%%%%%%%%%%%%%%%%%%%%%%%%%%%%%%%%%%%%%%%%%%%%%%%%%%%%%%%%%%%%%%%%%%%%%%%%
Longitudinal core-temperature measurements provide a particularly 
stringent test of thermal models because they resolve the dynamics 
throughout the entire swim instead of reporting only the initial and final states.
We test the models against four independent longitudinal datasets that
span warm and cold water, wetsuit and non-wetsuit conditions, and
individual and population-level responses.
The warm-water data include eight wetsuit swims at
\SI{24.5}{\degreeCelsius} \cite{Morton2025}
and twelve elite non-wetsuit swims at
\SI{25.5}{\degreeCelsius} \cite{Markey2025}.
The cold-water data include a prolonged non-wetsuit swim at
\SI{13.5}{\degreeCelsius} \cite{Roxburgh2026}
and wetsuit swims at \SI{10}{\degreeCelsius} \cite{Melau2019}.
Together, these data span {\it{near thermal equilibrium}}, {\it{transient heating}},
{\it{thermal overshoot}}, and {\it{delayed cooling}}.
For each trajectory, we identify the model parameters from the complete time series: metabolic heat production $Q$ and environmental conductance $H$ for models I and II, 
and, additionally, core--peripheral conductance $G$ and core thermal fraction $\nu_{\rm c}$ for model II, see Appendix for details. Importantly, we confront each model with the {\it{complete thermal trajectory}} of each swim, rather than with initial and final temperatures alone.
\vspace*{0.2cm}
%%%%%%%%%%%%%%%%%%%%%%%%%%%%%%%%%%%%%%%%%%%%%%%%%%%%%%%%%%%%%%%%%%%%%%%%%
\paragraph*{Warm-water swimming.}
%%%%%%%%%%%%%%%%%%%%%%%%%%%%%%%%%%%%%%%%%%%%%%%%%%%%%%%%%%%%%%%%%%%%%%%%%
Figure~\ref{fig04} tests the models on the heating side of the
thermal phase diagram.
The wetsuit swims at \SI{24.5}{\degreeCelsius} \cite{Morton2025} remain close to thermal
equilibrium over 80~min (left).
Both models reproduce this response, with {\sf{RMSE}}s of 0.06 and
\SI{0.04}{\degreeCelsius} for models I and II.
The simple energy balance (\ref{eq:model1}) captures the dominant thermal
behavior when core temperature approaches a nearly stationary state.
The individual non-wetsuit trajectories at
\SI{25.5}{\degreeCelsius} \cite{Markey2025} reveal dynamics that the one-compartment
model cannot represent (middle).
Core temperature initially rises and subsequently relaxes, and the
magnitude and time scale differ substantially among swimmers.
Model II captures these personalized trajectories with a median {\sf{RMSE}}
of \SI{0.06}{\degreeCelsius}.
The population mean exposes the same transient response particularly
clearly (right).
Model I predicts a monotonic approach toward equilibrium and therefore
misses the early temperature peak.
Model II reproduces the peak and subsequent relaxation and reduces the
{\sf{RMSE}} from 0.28 to \SI{0.10}{\degreeCelsius}.
This comparison identifies the physical role of the core--peripheral
description (\ref{eq:model2}):
A single thermal compartment predicts only {\it{monotonic exponential
dynamics}}, whereas two coupled thermal compartments introduce the
{\it{time-scale separation}} required for transient heating and relaxation.
The longitudinal data therefore test more than the net temperature
change: they distinguish thermal models that could appear similar from
pre- and post-swim measurements alone.
\vspace*{0.2cm}
%%%%%%%%%%%%%%%%%%%%%%%%%%%%%%%%%%%%%%%%%%%%%%%%%%%%%%%%%%%%%%%%%%%%%%%%%
\paragraph*{Cold-water swimming.}
%%%%%%%%%%%%%%%%%%%%%%%%%%%%%%%%%%%%%%%%%%%%%%%%%%%%%%%%%%%%%%%%%%%%%%%%%
Figure~\ref{fig05} tests the same physics on the cooling side of the
thermal phase diagram.
During the non-wetsuit swim in \SI{13.5}{\degreeCelsius} \cite{Roxburgh2026},
core temperature decreases from approximately
\SI{37.9}{\degreeCelsius} to below \SI{35}{\degreeCelsius} within the
first hour (left).
Model I captures this dominant cooling trend with an {\sf{RMSE}} of
\SI{0.06}{\degreeCelsius}.
Model II resolves additional early-time structure but yields a larger
{\sf{RMSE}} of \SI{0.15}{\degreeCelsius}.
This result shows that the additional physiological complexity of
model II does not automatically improve prediction when a trajectory
closely follows the monotonic dynamics of the fundamental energy
balance.
The wetsuit swims at \SI{10}{\degreeCelsius} \cite{Melau2019} reveal a fundamentally
different response.
Despite the colder water, several swimmers initially maintain or
slightly increase their core temperature before sustained cooling
begins.
Model I captures the overall cooling direction but cannot reproduce
this initial rise or plateau and yields a median {\sf{RMSE}} of
\SI{0.21}{\degreeCelsius} (middle).
Model II captures the initial dynamics, subsequent curvature, and
large swimmer-to-swimmer differences and reduces the median {\sf{RMSE}} to
\SI{0.06}{\degreeCelsius} (right).
The predicted trajectories also reproduce the markedly different
times at which individual swimmers cross the mild-hypothermia and
hypothermia boundaries.
%%%%%%%%%%%%%%%%%%%%%%%%%%%%%%%%%%%%%%%%%%%%%%%%%%%%%%%%%%%%%%%%%%%%%%%%%
\begin{figure*}[t]
\centering
\includegraphics[width=\textwidth]{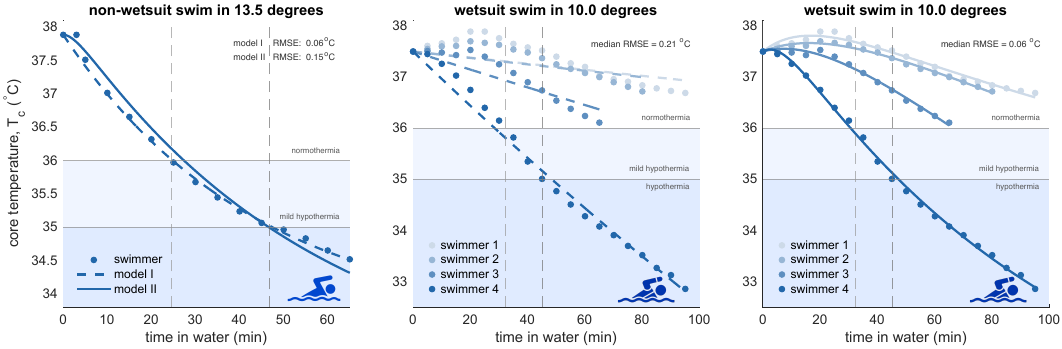}
\caption{\textbf{Core temperature dynamics during cold-water swim.}
Dots show the temperature profile of a 
non-wetsuit swim in 13 to \SI{14}{\degreeCelsius} \cite{Roxburgh2026}; 
dashed and solid curves show models I and II 
with {\sf{RMSE}} of 0.06 and \SI{0.15}{\degreeCelsius} (left).
Dots show n=4  
representative wetsuit swims in \SI{10.0}{\degreeCelsius} \cite{Melau2019};
dashed curves show that models I 
captures the overall cooling trend but
cannot represent the early temperature dynamics 
with a median {\sf{RMSE}} of \SI{0.21}{\degreeCelsius} (middle).
Dots show the same n=4  
representative wetsuit swims in \SI{10.0}{\degreeCelsius} \cite{Melau2019};
solid curves show that model II
captures the initial slight rise, plateau, and swimmer-specific cooling 
with a median {\sf{RMSE}} of \SI{0.06}{\degreeCelsius} (right).
Horizontal solid lines mark mild hypothermia at \SI{36.0}{\degreeCelsius} and
hypothermia at \SI{35.0}{\degreeCelsius};
vertical dashed lines mark crossings of the mild-hypothermia and hypothermia thresholds.}
\label{fig05}
\end{figure*}
%%%%%%%%%%%%%%%%%%%%%%%%%%%%%%%%%%%%%%%%%%%%%%%%%%%%%%%%%%%%%%%%%%%%%%%%%

\vspace*{0.2cm}
\noindent Warm- and cold-water calibrations together reveal a consistent
hierarchy.
Model I captures the dominant direction and time scale of thermal
drift when one thermal mode governs the response.
Model II becomes necessary when core--peripheral heat exchange creates
multiple time scales, including overshoot, plateau, delayed cooling,
and subsequent relaxation.
Since we evaluate these dynamics throughout each swim, the
longitudinal trajectories directly probe this distinction between the
two models.
\vspace*{0.2cm}
%%%%%%%%%%%%%%%%%%%%%%%%%%%%%%%%%%%%%%%%%%%%%%%%%%%%%%%%%%%%%%%%%%%%%%%%%
\paragraph*{Implications.}
%%%%%%%%%%%%%%%%%%%%%%%%%%%%%%%%%%%%%%%%%%%%%%%%%%%%%%%%%%%%%%%%%%%%%%%%%
The theory reveals three conclusions: 
First, open-water thermalbalance follows a competition between metabolic heat production and environmental heat transfer rather than water temperature alone. The critical water temperature scaling law in equation~(\ref{eq:critical}) provides the simplest expression of this balance. 
This equilibrium boundary does not define a physiological safety limit, and its application to actual swims assumes constant metabolic heat production and effective conductance.
Second, body size matters through mass and surface area; mass sets heat capacity and heat generation, while surface area controls environmental exchange. This distinction explains why body mass index alone cannot fully characterize thermal response. 
Third, the time course contains information that endpoint measurements ignore. The two-compartment model II (\ref{eq:model2}) predicts plateaus, overshoots, curvature, and delayed cooling that the single-compartment model I (\ref{eq:model1}) cannot reproduce. 
Calibration with complete longitudinal trajectories probes substantially more of the governing dynamics than comparisons of pre- and post-swim temperatures alone, although it does not establish predictive accuracy on unseen data. Yet, the simple physics-based model I remains valuable because it exposes the governing scaling law with almost no physiological complexity.
\vspace*{0.2cm}
%%%%%%%%%%%%%%%%%%%%%%%%%%%%%%%%%%%%%%%%%%%%%%%%%%%%%%%%%%%%%%%%%%%%%%%%%
\paragraph*{Limitations.}
%%%%%%%%%%%%%%%%%%%%%%%%%%%%%%%%%%%%%%%%%%%%%%%%%%%%%%%%%%%%%%%%%%%%%%%%%
First, long-duration core temperature data are rare and predominantly involve trained or elite swimmers who sustain high metabolic heat production; broader validation across recreational swimmers, swim intensities, and body compositions will further establish swimmer-specific thermal boundaries.
Second, acclimatization, fueling, wind, sun, stroke mechanics, and wetsuit 
composition introduce additional sources of individualization that our current 
personalized model does not yet resolve.
Third, we interpret conventional core-temperature bands as thermal 
states rather than universal physiological limits, since individual 
tolerance can vary substantially.
These factors provide natural extensions of the personalized framework 
rather than limitations of the underlying thermal scaling law.
\vspace*{0.2cm}
%%%%%%%%%%%%%%%%%%%%%%%%%%%%%%%%%%%%%%%%%%%%%%%%%%%%%%%%%%%%%%%%%%%%%%%%%
\paragraph*{Conclusion.}
%%%%%%%%%%%%%%%%%%%%%%%%%%%%%%%%%%%%%%%%%%%%%%%%%%%%%%%%%%%%%%%%%%%%%%%%%
The phase-boundary law $T_w^{\rm crit}=T_{\rm c}^{0}-Q/H$ separates heating from cooling in open water, but no single water temperature can separate safe from unsafe swimming for every athlete. The same balance of heat production and heat loss that keeps an emperor penguin warm in near-freezing Antarctic water governs the thermal 
boundary of a human swimmer: the critical temperature depends on body geometry, metabolic heat production, and insulation, while exposure time also depends on transient core--periphery dynamics. Models I--III organize these effects from physics to physiology to personalization. Across independent warm- and cold-water field data, full-trajectory calibration reproduces both directions of thermal drift, transient plateaus and peaks, and swimmer-to-swimmer variability that endpoint comparisons cannot reveal. This physics-based view turns water temperature from a universal cutoff into one coordinate of a swimmer-specific thermal phase diagram.
%%%%%%%%%%%%%%%%%%%%%%%%%%%%%%%%%%%%%%%%%%%%%%%%%%%%%%%%%%%%%%%%%%%%%%%%%
\vspace*{0.4cm}
%%%%%%%%%%%%%%%%%%%%%%%%%%%%%%%%%%%%%%%%%%%%%%%%%%%%%%%%%%%%%%%%%%%%%%%%%
%\paragraph*{Appendix.}
\begin{center}
{\bf{Appendix}}
\end{center}
%\vspace*{0.2cm}
%%%%%%%%%%%%%%%%%%%%%%%%%%%%%%%%%%%%%%%%%%%%%%%%%%%%%%%%%%%%%%%%%%%%%%%%%
This appendix provides the derivation, parameterization, and numerical implementation of models I--III introduced in the main text. We first derive the single-compartment thermal scaling law, then formulate the core--peripheral dynamics as a linear two-state system, and finally describe the swimmer-specific scalings for body size, swim pace, and insulation. We conclude with an illustrative Level-II thermal exposure map that combines the phase boundary with the time required to reach conventional core-temperature states.
\vspace*{0.2cm}
%%%%%%%%%%%%%%%%%%%%%%%%%%%%%%%%%%%%%%%%%%%%%%%%%%%%%%%%%%%%%%%%%%%%%%%%%
\paragraph*{Model I: Single-compartment thermal scaling.}
%%%%%%%%%%%%%%%%%%%%%%%%%%%%%%%%%%%%%%%%%%%%%%%%%%%%%%%%%%%%%%%%%%%%%%%%%
We represent the swimmer as a single well-mixed thermal compartment with core temperature $T_{\rm c}$, body mass $m$, and effective specific heat $c$. Conservation of energy gives
\begin{equation}
mc\,\dot T_{\rm c}=Q-H(T_{\rm c}-T_{\rm w}),
\label{eq:S1}
\end{equation}
where $Q$ denotes metabolic heat production, $T_{\rm w}$ the water temperature, and $H$ the whole-body environmental thermal conductance. We write
\begin{equation}
H=h_{\rm eff}A,
\label{eq:S2}
\end{equation}
with effective heat-transfer coefficient $h_{\rm eff}$ and body surface area $A$. The calculations use the Du Bois relation
\begin{equation}
A=0.007184\,m^{0.425}h^{0.725},
\label{eq:S3}
\end{equation}
where $m$ is measured in kilograms and height $h$ in centimeters.
For constant $Q$, $H$, and $T_{\rm w}$, Eq.~(\ref{eq:S1}) has the closed-form solution
\begin{equation}
T_{\rm c}(t)=T_\infty+
\left(T_{\rm c}^{0}-T_\infty\right)e^{-t/\tau},
\label{eq:S4}
\end{equation}
with
\begin{equation}
T_\infty=T_{\rm w}+\frac{Q}{H}
\quad \mbox{and} \quad
\tau=\frac{mc}{H}.
\label{eq:S5}
\end{equation}
These two quantities have distinct physical meanings. The steady-state offset $T_\infty-T_{\rm w}=Q/H$ follows from the competition between metabolic heat production and environmental heat exchange, whereas the time scale $\tau$ follows from the ratio between thermal capacitance $mc$ and conductance $H$. Thus $mc$ controls how rapidly the swimmer approaches steady state but does not alter the steady state itself.
The initial direction of thermal drift follows directly from Eq.~(\ref{eq:S1}),
\begin{equation}
\dot T_{\rm c}(0)=\frac{1}{mc}
\left[Q-H(T_{\rm c}^{0}-T_{\rm w})\right]\,.
\label{eq:S6}
\end{equation}
Thermal neutrality requires $\dot T_{\rm c}(0)=0$, or equivalently $T_\infty=T_{\rm c}^{0}$. This condition defines the critical-water-temperature scaling law
\begin{equation}
%\boxed{
T_{\rm w}^{\rm crit}=T_{\rm c}^{0}-\frac{Q}{H}\,.
%}.
\label{eq:S7}
\end{equation}
For $T_{\rm w}<T_{\rm w}^{\rm crit}$ the model predicts net cooling; for $T_{\rm w}>T_{\rm w}^{\rm crit}$ it predicts net heating. The dimensionless thermal driving parameter
\begin{equation}
\Pi=\frac{Q}{H(T_{\rm c}^{0}-T_{\rm w})}
\label{eq:S8}
\end{equation}
provides an equivalent representation: $\Pi<1$ denotes cooling, $\Pi=1$ thermal neutrality, and $\Pi>1$ heating.
\vspace*{0.2cm}
%%%%%%%%%%%%%%%%%%%%%%%%%%%%%%%%%%%%%%%%%%%%%%%%%%%%%%%%%%%%%%%%%%%%%%%%%
\paragraph*{Model II: Core--peripheral thermal dynamics.}
%%%%%%%%%%%%%%%%%%%%%%%%%%%%%%%%%%%%%%%%%%%%%%%%%%%%%%%%%%%%%%%%%%%%%%%%%
A single thermal compartment contains one thermal time scale and therefore produces a monotonic exponential trajectory. To resolve plateaus, overshoots, delayed cooling, and relaxation, we divide the swimmer into coupled core and peripheral compartments,
\begin{equation}
\begin{array}{llll}
C_{\rm c} & \dot T_{\rm c} &= \, Q \, - & G\,(T_{\rm c}-T_{\rm p}), \\
C_{\rm p} & \dot T_{\rm p} &=           & G\,(T_{\rm c}-T_{\rm p})-H(T_{\rm p}-T_{\rm w}),
\end{array}
\label{eq:S10}
\end{equation}
Here $T_{\rm p}$ denotes peripheral temperature, $G$ the core--peripheral conductance, and
\begin{equation}
C_{\rm c}=\nu_{\rm c}mc,
\quad \mbox{and} \quad
C_{\rm p}=(1-\nu_{\rm c})mc
\label{eq:S11}
\end{equation}
are the core and peripheral thermal capacitances. The parameter $\nu_{\rm c}\in(0,1)$ partitions the total thermal capacitance $mc$ between core and periphery. 
Equations~(\ref{eq:S10}) preserve the total energy balance. Adding both equations eliminates the internal exchange term $G(T_{\rm c}-T_{\rm p})$ and gives
\begin{equation}
C_{\rm c}\dot T_{\rm c}+C_{\rm p}\dot T_{\rm p}
=Q-H(T_{\rm p}-T_{\rm w}).
\label{eq:S12}
\end{equation}
Thus the core-peripheral conductance $G$ redistributes heat internally but neither creates nor removes energy.
For constant coefficients, we can summarize the model as
\begin{equation}
\dot{\bm T}=\bm M\bm T+\bm b
\quad \mbox{with} \quad
\bm T=
\begin{bmatrix}T_{\rm c}\\T_{\rm p}\end{bmatrix}
\label{eq:S13}
\end{equation}
where
\begin{equation}
\bm M=
\begin{bmatrix}
-G/C_{\rm c} & G/C_{\rm c}\\
G/C_{\rm p} & -(G+H)/C_{\rm p}
\end{bmatrix}
\;\, \mbox{and} \;\,
\bm b=
\begin{bmatrix}
Q/C_{\rm c}\\HT_{\rm w}/C_{\rm p}
\end{bmatrix}.
\label{eq:S14}
\end{equation}
The exact solution is
\begin{equation}
\bm T(t)=\bm T_\infty+
\exp(\bm M t)\left(\bm T^0-\bm T_\infty\right)
\, \mbox{with} \,
\bm T_\infty=-\bm M^{-1}\bm b.
\label{eq:S15}
\end{equation}
The two eigenvalues of $\bm M$ define two thermal time scales. Their separation allows the peripheral compartment to respond rapidly to the water while the core responds more slowly, which produces the non-monotonic core-temperature dynamics observed in longitudinal swims.
At steady state, Eq.~(\ref{eq:S10}.1) gives $Q=G(T_{\rm c,\infty}-T_{\rm p,\infty})$, while Eq.~(\ref{eq:S10}.2) gives $Q=H(T_{\rm p,\infty}-T_{\rm w})$. Hence
\begin{equation}
T_{\rm p,\infty}=T_{\rm w}+\frac{Q}{H}
\quad \mbox{and} \quad
T_{\rm c,\infty}=T_{\rm w}+\frac{Q}{H}+\frac{Q}{G}.
\label{eq:S16}
\end{equation}
Model II therefore modifies the transient core response and introduces an internal core--peripheral temperature difference. The environmental heat rejection at steady state remains $Q$.\\[6.pt]
\noindent For the longitudinal swim data, model I identifies $Q$ and $H$ from each complete core-temperature trajectory. Model II identifies $Q$, $G$, $H$, and $\nu_{\rm c}$ and, where required by the available data, the initial peripheral temperature $T_{\rm p}^{0}$. Parameters minimize the sum of squared errors over the time series; reported {\sf{RMSE}} values use the same complete trajectories. This procedure deliberately uses the temporal structure of the swim rather than only the initial and final core temperatures.
\vspace*{0.2cm}
%%%%%%%%%%%%%%%%%%%%%%%%%%%%%%%%%%%%%%%%%%%%%%%%%%%%%%%%%%%%%%%%%%%%%%%%%
\paragraph*{Model III: Personalized thermal boundaries.}
%%%%%%%%%%%%%%%%%%%%%%%%%%%%%%%%%%%%%%%%%%%%%%%%%%%%%%%%%%%%%%%%%%%%%%%%%
Model III retains the phase-boundary structure of Eq.~(\ref{eq:S7}) and allows swimmer characteristics and swimming conditions to adjust. The main text considers body size, swim intensity, and insulation separately.\\[6.pt]
\noindent{\emph{Body size.}} Body size enters through heat-producing mass and heat-losing surface area. With Eq.~(\ref{eq:S3}),
\begin{equation}
T_{\rm w}^{\rm crit}(m,h)
=T_{\rm c}^{0}-
\frac{Q}{h_{\rm eff}\,0.007184\,m^{0.425}h^{0.725}}.
\label{eq:S17}
\end{equation}
If metabolic heat production scales approximately with mass, $Q=qm$, then
\begin{equation}
\frac{Q}{H}
=\frac{q}{0.007184\,h_{\rm eff}}
\,m^{0.575}h^{-0.725}.
\label{eq:S18}
\end{equation}
The exponent $0.575=1-0.425$ makes the mass-to-area scaling explicit: at fixed height and mass-specific heat production, larger mass increases the metabolic temperature offset and lowers the critical water temperature.\\[6.pt]
\noindent {\emph{Swim intensity.}}
We represent swim intensity through velocity $v$ and use the effective mass-specific heat-production relation
\begin{equation}
Q(v,m)=m q(v)
\quad \mbox{with} \quad
q(v)=q_{\rm ref}
\left(\frac{v}{v_{\rm ref}}\right)^n,
\label{eq:S19}
\end{equation}
The middle panel of Fig.~2 uses $v_{\rm ref}=\SI{1.0}{m\,s^{-1}}$ and $n=1.5$. Rather than equating $q(v)$ to gross metabolic power, we calibrate the thermal offset $Q/H$ to the range identified from the longitudinal warm- and cold-water calibrations. The implementation uses the 85th-percentile of the calibrated $Q/H$ values as a conservative reference offset and then applies Eq.~(\ref{eq:S19}) to vary pace. This construction anchors the phase-boundary magnitude to observed thermal trajectories while retaining a nonlinear intensity dependence.
%%%%%%%%%%%%%%%%%%%%%%%%%%%%%%%%%%%%%%%%%%%%%%%%%%%%%%%%%%%%%%%%%%%%%%%%%
\begin{figure*}[t]
\centering
\includegraphics[width=\textwidth]{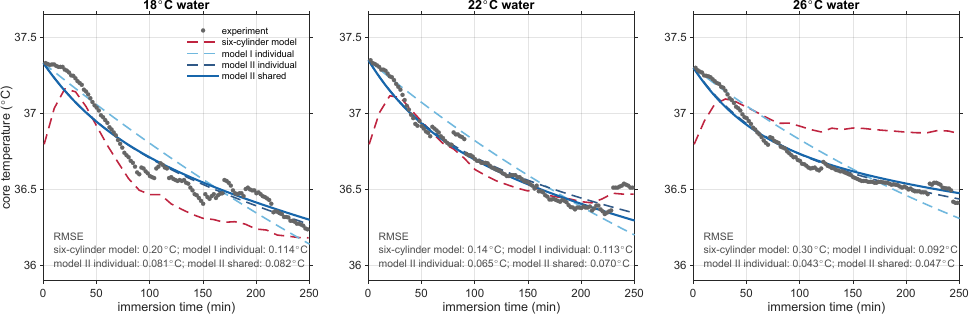}
\caption{\textbf{Cold-water immersion data and comparison with six-cylinder thermoregulation model.} Core temperature trajectories at 18$^\circ$C, 22$^\circ$C, and 26$^\circ$C. 
Black dots show experimental measurements \cite{Castellani2023};  
dashed red curves represent predictions of six-cylinder thermoregulation model \cite{Castellani2023};
dashed light and dark blue curves show independent calibrations of models I and II, and 
solid dark blue curves represent model II with shared parameters and temperature-dependent conductance $H(T_{\rm w})$.}
\label{fig06}
\end{figure*}
%%%%%%%%%%%%%%%%%%%%%%%%%%%%%%%%%%%%%%%%%%%%%%%%%%%%%%%%%%%%%%%%%%%%%%%%%
Because swim pace $p$ in minutes per 100~m satisfies $v=100/(60p)$, we
can rewrite Eq.~(\ref{eq:S19}) as
\begin{equation}
q(p)=q_{\rm ref}
\left(\frac{100}{60 \, p\,v_{\rm ref}}\right)^n.
\label{eq:S20}
\end{equation}
Faster pace therefore increases $Q/H$ and lowers $T_{\rm w}^{\rm crit}$.\\[6.pt]
\noindent {\emph{Insulation.}}
We describe insulation through the relative environmental conductance
\begin{equation}
\eta=\frac{H}{H_0},
\label{eq:S21}
\end{equation}
where $H_0$ denotes a reference conductance. The critical temperature becomes
\begin{equation}
T_{\rm w}^{\rm crit}(\eta)
=T_{\rm c}^{0}-\frac{Q}{\eta H_0}.
\label{eq:S22}
\end{equation}
Greater insulation corresponds to $\eta<1$. It reduces environmental heat loss, increases the metabolic temperature offset $Q/H$, and lowers the critical water temperature. This parameterization does not assign a universal $\eta$ to model the skin or wetsuit; instead, it  isolates the thermal consequence of changing effective conductance.
\vspace*{0.2cm}
%%%%%%%%%%%%%%%%%%%%%%%%%%%%%%%%%%%%%%%%%%%%%%%%%%%%%%%%%%%%%%%%%%%%%%%%%
\paragraph*{Reference parameters and sensitivity analysis.}
%%%%%%%%%%%%%%%%%%%%%%%%%%%%%%%%%%%%%%%%%%%%%%%%%%%%%%%%%%%%%%%%%%%%%%%%%
For Figure \ref{fig01}, we use
$m=\SI{70}{kg}$, $h=\SI{175}{cm}$,
$c=\SI{3470}{J/kg\,K^{-1}}$, and
$h_{\rm eff}=\SI{8.11}{W/m^{2}\,K^{-1}}$.
We select the reference metabolic heat production such that
the reference swimmer has
$T_{\rm w}^{\rm crit}=\SI{26.5}{\degreeCelsius}$.
This choice sets the illustrative phase boundary in
Figure \ref{fig01}.
For Figure \ref{fig02}, we retain the same reference swimmer
and conductance but increase the reference metabolic heat
production to $Q_0=\SI{337}{W}$, which corresponds to
$T_{\rm w}^{\rm crit}=\SI{15.0}{\degreeCelsius}$.
We vary body mass from 50 to 100 kg, height from
150 to 200 cm, swim pace from 1:15 to 2:30 min/100 m
with a velocity exponent of $n=1.5$, and relative
thermal conductance from $H/H_0=0.7$ to $1.3$.
At fixed heat production, the latter range changes
the critical water temperature from
\SI{5.4}{\degreeCelsius} to
\SI{20.2}{\degreeCelsius}, compared with
\SI{15.0}{\degreeCelsius} at baseline.
For the swim-intensity analysis, we use a separately
calibrated conductance anchored to the longitudinal
temperature data.
To assess parameter sensitivity, we additionally consider
illustrative variations of $\pm20\%$ in reference metabolic
heat production and $n=1.0$ to $n=2.0$ in the velocity
exponent. 
Varying heat production from \SI{270}{W}
to \SI{404}{W} shifts $T_{\rm w}^{\rm crit}$ from
\SI{19.5}{\degreeCelsius} to
\SI{10.5}{\degreeCelsius} at fixed conductance.
Varying the velocity exponent from $n=1.0$ to
$n=2.0$, relative to the baseline $n=1.5$,
changes the critical temperature by
$-2.2$ to $+1.8\,^\circ\mathrm{C}$ at
2:30 min/100 m, and by
$+3.7$ to $-4.3\,^\circ\mathrm{C}$ at
1:15 min/100 m.
These illustrative reference calculations differ from
the trajectory-specific parameter identification for
the longitudinal temperature data in Figures
\ref{fig04} and \ref{fig05}.
The effective conductance incorporates tissue and clothing
insulation, peripheral circulation, and environmental
heat exchange; it is not the water-side convective
coefficient alone and may vary with swimming conditions.
\vspace*{0.2cm}
%%%%%%%%%%%%%%%%%%%%%%%%%%%%%%%%%%%%%%%%%%%%%%%%%%%%%%%%%%%%%%%%%%%%%%%%%
\paragraph*{Experimental calibration and model comparison.}
%%%%%%%%%%%%%%%%%%%%%%%%%%%%%%%%%%%%%%%%%%%%%%%%%%%%%%%%%%%%%%%%%%%%%%%%%
Earlier studies developed thermoregulation models for water immersion and prolonged cold exposure, ranging from heat-transfer and compartmental formulations \cite{Miller1974,Tikuisis1988,Tarlochan2005,Xu2005} to geometrically detailed three-dimensional models \cite{Castellani2021}. The Cold Weather Ensemble Decision Aid further demonstrates the application of thermoregulation models to cold-exposure risk assessment \cite{Xu2021}. These approaches resolve physiological and spatial mechanisms beyond the scope of our reduced-order formulation. Our model instead represents their combined effects through effective thermal conductances and heat capacities.\\
\quad Our core--peripheral model follows the classical two-node approach to human thermoregulation \cite{Gagge1971}. The internal conductance $G$ represents effective heat exchange between core and peripheral tissues through conduction and blood perfusion. Peripheral vasoconstriction, changes in blood flow, and thermoregulatory responses modify this conductance, while shivering can increase metabolic heat production $Q$ during prolonged cold exposure. Our model captures these mechanisms through constant effective parameters rather than explicit physiological regulation.\\
\quad To assess this simplification, we compare models I and II with reported water-immersion experiments and with a six-cylinder thermoregulation model \cite{Castellani2023}. Figure \ref{fig06} compares experimental core-temperature trajectories at water temperatures of 18, 22, and 26$^\circ$C with the published six-cylinder model, our reduced-order models I and II calibrated with individual parameterizations for each temperature, and model II with a shared parameterization and a temperature-dependent external conductance.
We report the root-mean-square error between 
the computationally modeled temperature $T_{\mathrm{c}}^{\mathrm{model}}$ and
the experimentally measured temperatures $T_{\mathrm{c}}^{\mathrm{exp}}$,
\begin{equation}
{\mathsf{RMSE}}
=
\sqrt{
{1}/{n} \;
\mbox{$\sum_{i=1}^{n}$} 
[\,
T_{\mathrm{c}}^{\mathrm{model}}(t_i)
-
T_{\mathrm{c}}^{\mathrm{data}}(t_i)
\,]^2 
} \,,
\end{equation}
where $n$ is the number of experimental time points.
Because we estimate the model parameters from the complete experimental trajectories, the error quantifies goodness of fit rather than independent predictive accuracy.
The root mean squared error 
for the six-cylinder thermoregulation model varies 
from 0.14$^\circ$C to 0.30$^\circ$C. 
For independent parameterizations of models I and II, it varies 
from 0.092$^\circ$C to 0.114$^\circ$C and  
from 0.042$^\circ$C to 0.081$^\circ$C.
For the shared parameterization, model II with a
temperature-dependent external conductance achieves similar errors
from 0.047$^\circ$C to 0.082$^\circ$C. 
While these results show that a reduced-order core--peripheral model reproduces the observed thermal dynamics under cold-water immersion, they do not establish the uniqueness of the calibrated physiological parameters or predictive accuracy outside the fitted trajectories.
\vspace*{0.2cm}
%%%%%%%%%%%%%%%%%%%%%%%%%%%%%%%%%%%%%%%%%%%%%%%%%%%%%%%%%%%%%%%%%%%%%%%%
\paragraph*{Parameter interpretation and identifiability.}
%%%%%%%%%%%%%%%%%%%%%%%%%%%%%%%%%%%%%%%%%%%%%%%%%%%%%%%%%%%%%%%%%%%%%%%%%
The calibrated core--peripheral conductance $G$ ranges from \SI{36.7}{W/K} to \SI{150.0}{W/K} for the four cold-water swimmers, with a median of \SI{53.5}{W/K}. The corresponding core heat-capacity fractions $\nu_{\rm{c}}$ range from 0.419 to 0.506, with a median of 0.462. In contrast, the warm-water calibrations yield broader parameter ranges, and $\nu_{\rm{c}}$ reaches its imposed lower bound of 0.05 for 10 of the 12 individual trajectories. These boundary solutions indicate limited practical identifiability of the individual parameters. Classical two-node thermoregulation models assign approximately 90\% of body mass to the core and
represent core--skin heat transfer through tissue conductance and blood perfusion. Our parameters instead represent effective thermal capacitance and conductance and therefore need not coincide with anatomical compartment fractions or tissue conductance alone. Because only core temperature is measured, the available trajectories do not uniquely constrain the internal heat-transfer mechanism. The improved agreement of model II thus supports the usefulness of a two-compartment
description but does not independently establish the proposed physiological mechanism.
\vspace*{0.2cm}
%%%%%%%%%%%%%%%%%%%%%%%%%%%%%%%%%%%%%%%%%%%%%%%%%%%%%%%%%%%%%%%%%%%%%%%%%
\paragraph*{Longitudinal parameter identification.}
%%%%%%%%%%%%%%%%%%%%%%%%%%%%%%%%%%%%%%%%%%%%%%%%%%%%%%%%%%%%%%%%%%%%%%%%%
The calibration uses independent warm- and cold-water trajectories described in the main text. For model I, the parameter vector is
\begin{equation}
\bm p_{\rm I}=[\,Q,H\,].
\label{eq:S23}
\end{equation}
For model II, the parameter vector is
\begin{equation}
\bm p_{\rm II}=[\,Q,G,H,\nu_{\rm c},T_{\rm p}^{0}\,],
\label{eq:S24}
\end{equation}
when $T_{\rm p}^{0}$ is not independently prescribed. In data sets with a reported or imposed initial peripheral temperature, the corresponding parameter is fixed and omitted from the optimization. We constrain all conductances and heat production to positive values and constrain $0<\nu_{\rm c}<1$. Numerical optimization minimizes
\begin{equation}
\mathcal{L}(\bm p)
=\mbox{$\sum_{i=1}^{n}$}
\left[ \, T_{\rm c}^{\rm model}(t_i;\bm p)-T_{\rm c}^{\rm data}(t_i) \, \right]^2,
\label{eq:S25}
\end{equation}
and reports the root mean squared error 
\begin{equation}
{\mathsf{RMSE}}=\sqrt{{1}/{n} \; \mathcal{L}} \, .
\label{eq:S26}
\end{equation}
\vspace*{0.2cm}
%%%%%%%%%%%%%%%%%%%%%%%%%%%%%%%%%%%%%%%%%%%%%%%%%%%%%%%%%%%%%%%%%%%%%%%%%
\paragraph*{Core-temperature threshold crossings.}
%%%%%%%%%%%%%%%%%%%%%%%%%%%%%%%%%%%%%%%%%%%%%%%%%%%%%%%%%%%%%%%%%%%%%%%%%
To quantify the temporal agreement between the calibrated models and the experimental trajectories, we determine the first downward crossings of core temperatures of 36$^\circ$C and 35$^\circ$C. 
For the swimmer at 13.5$^\circ$C in Figure \ref{fig05}, left, the digitized trajectory crosses these thresholds after 24.6 and 46.8 min \cite{Roxburgh2026}. 
Model I predicts crossing times of 24.9 and 47.5 min and model II predicts 24.5 and 46.4 min.
For the coldest swimmer in the 10$^\circ$C cohort, swimmer 4 in Figure \ref{fig05}, middle and right, the digitized trajectory crosses these thresholds after 32.3 and 45.2 min \cite{Melau2019}. The remaining three swimmers do not cross either threshold during their recorded swims. 
Model I predicts 28.5 and 48.4 min and model II predicts 30.6 and 47.3 min for swimmer 4.  
Thus, model II reproduces all four observed crossing times within approximately 2.1 min. 
While these comparisons do not establish independent predictive accuracy or individualized safe exposure durations, they demonstrate that the calibrated models accurately reproduce the observed core-temperature threshold-crossing times.
%%%%%%%%%%%%%%%%%%%%%%%%%%%%%%%%%%%%%%%%%%%%%%%%%%%%%%%%%%%%%%%%%%%%%%%%%
\begin{figure}[t]
\centering
\includegraphics[width=0.4\textwidth]{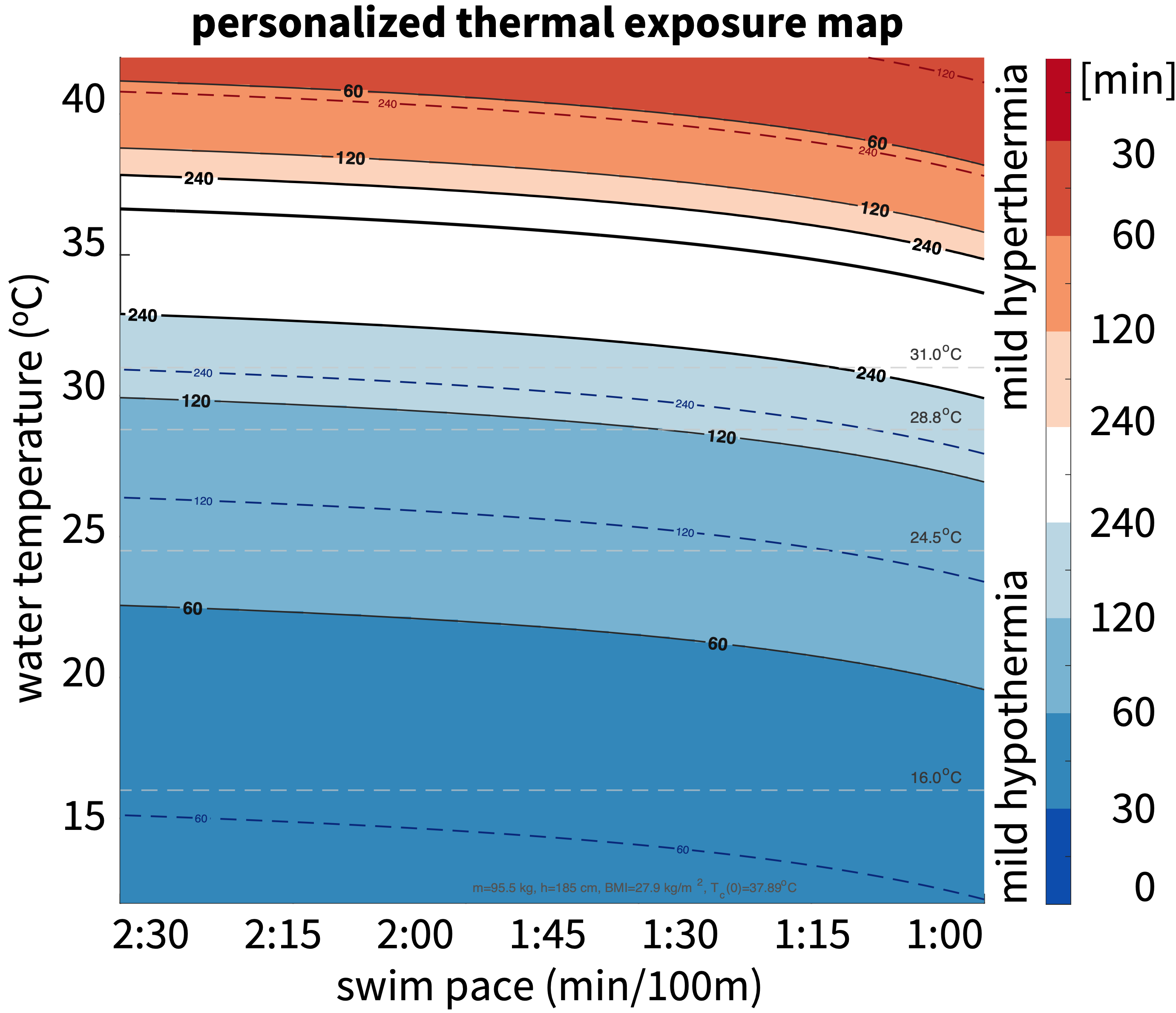} % roxburgh
\caption{\textbf{Personalized thermal exposure map for marathon swimmer.}
The marathon swimmer  
of Figure \ref{fig05} (left) in the main text has a
mass $\SI{95.5}{kg}$, 
height $\SI{185}{cm}$,
${\rm BMI}=27.9\,\si{kg/m^2}$, and 
initial core temperature $T_{\rm c}^{0}=\SI{37.89}{\degreeCelsius}$.
The calibration uses the non-wetsuit swim in \SI{13.5}{\degreeCelsius}
water, with an average pace near 1:36~min/100~m.
The map varies swim pace from 2:30 to 1:00~min/100~m while keeping body
size fixed.
Blue and red regions mark first-passage times to 
mild hypothermia at $T_{\rm c}=\SI{36.0}{\degreeCelsius}$ and 
mild hyperthermia at $T_{\rm c}=\SI{38.5}{\degreeCelsius}$.
Regions distinguish time to passage, from 30 to 240 minutes. 
Solid black contours mark the 
mild hypothermia and mild hyperthermia boundaries.
Dashed contours mark first-passage times to hypothermia at
$\SI{35.0}{\degreeCelsius}$ and hyperthermia at
$\SI{40.0}{\degreeCelsius}$.
Horizontal reference lines indicate 16.0, 24.5, 28.8, and
$\SI{31.0}{\degreeCelsius}$.}
\label{fig07}
\end{figure}
%%%%%%%%%%%%%%%%%%%%%%%%%%%%%%%%%%%%%%%%%%%%%%%%%%%%%%%%%%%%%%%%%%%%%%%%%
%%%%%%%%%%%%%%%%%%%%%%%%%%%%%%%%%%%%%%%%%%%%%%%%%%%%%%%%%%%%%%%%%%%%%%%%%
\begin{figure}[t]
\centering
\includegraphics[width=0.4\textwidth]{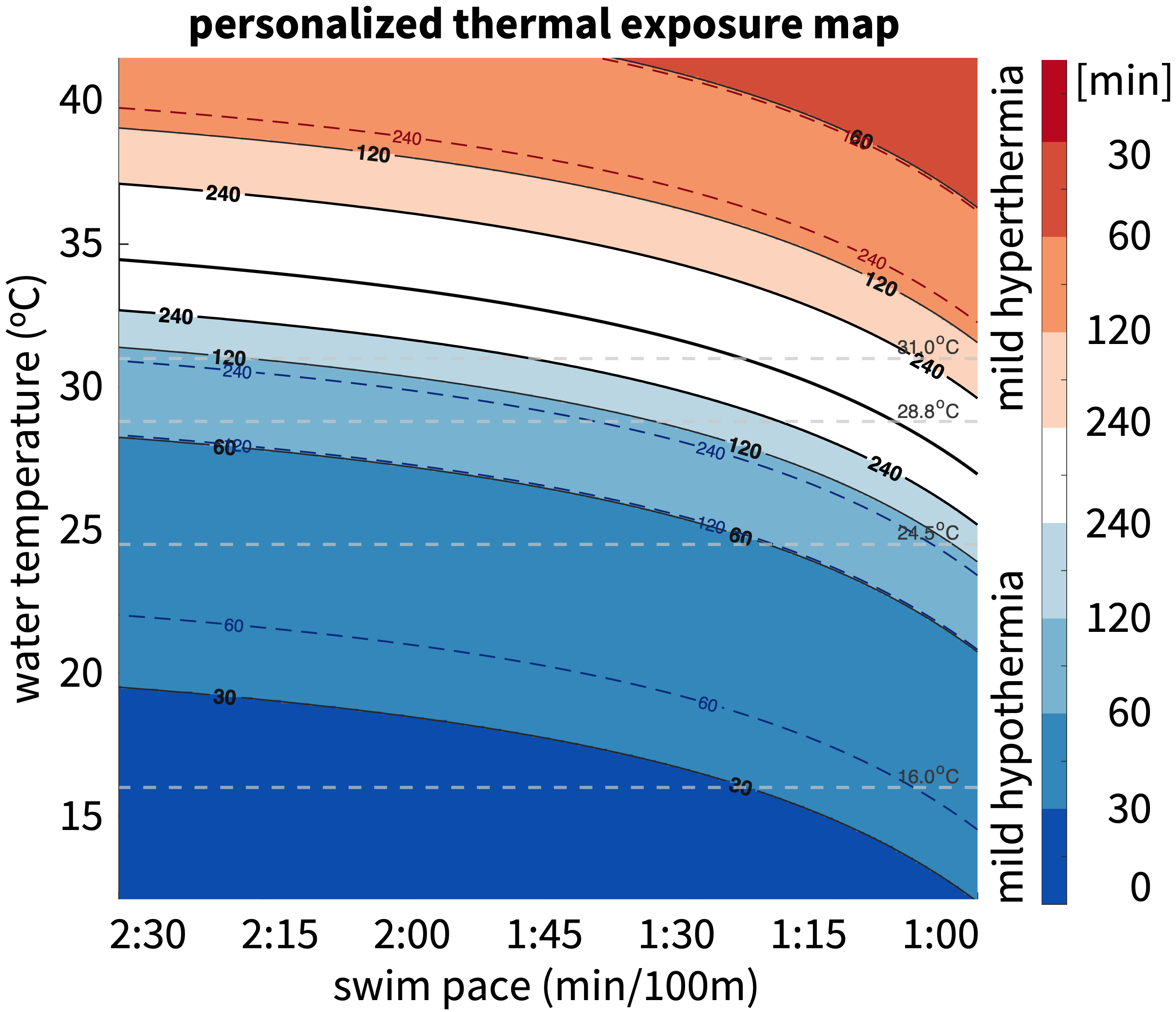} % henry
\caption{\textbf{Personalized thermal exposure map for recreational
swimmer.}
The recreational swimmer has a
mass $\SI{68}{kg}$, 
height $\SI{175}{cm}$,
${\rm BMI}=22.2\,\si{kg/m^2}$, 
initial core temperature $T_{\rm c}^{0}=\SI{37.0}{\degreeCelsius}$, 
and reference pace 1:50~min/100~m.
The calibration uses a core temperature of
$T_{\rm c}=\SI{36.0}{\degreeCelsius}$ after 30~min in
$\SI{18}{\degreeCelsius}$ water.
The map varies swim pace from 2:30 to 1:00~min/100~m while keeping body
size fixed.
Blue and red regions mark first-passage times to 
mild hypothermia at $T_{\rm c}=\SI{36.0}{\degreeCelsius}$ and 
mild hyperthermia at $T_{\rm c}=\SI{38.5}{\degreeCelsius}$.
Regions distinguish time to passage, from 30 to 240 minutes. 
Solid black contours mark the 
mild hypothermia and mild hyperthermia boundaries.
Dashed contours mark first-passage times to hypothermia at
$\SI{35.0}{\degreeCelsius}$ and hyperthermia at
$\SI{40.0}{\degreeCelsius}$.
Horizontal reference lines indicate 16.0, 24.5, 28.8, and
$\SI{31.0}{\degreeCelsius}$.}
\label{fig08}
\end{figure}
%%%%%%%%%%%%%%%%%%%%%%%%%%%%%%%%%%%%%%%%%%%%%%%%%%%%%%%%%%%%%%%%%%%%%%%%%
\vspace*{0.2cm}
%%%%%%%%%%%%%%%%%%%%%%%%%%%%%%%%%%%%%%%%%%%%%%%%%%%%%%%%%%%%%%%%%%%%%%%%%
\paragraph*{Model comparison and complexity.}
%%%%%%%%%%%%%%%%%%%%%%%%%%%%%%%%%%%%%%%%%%%%%%%%%%%%%%%%%%%%%%%%%%%%%%%%%
To assess whether the improved agreement of model II justifies its additional parameters, we compare both models with the Akaike information criterion, 
${\mathsf{AIC}}=n\ln(\mathrm{RSS}/n)+2k$, where $n$ denotes the number of observations, $k$ the number of calibrated parameters, and $\mathrm{RSS}$ the residual sum of squares. To account for the relatively small number of observations, particularly after temporal subsampling, we adopt the small-sample corrected criterion, 
${\mathsf{AICc}}
={\mathsf{AIC}}+2k(k+1)/(n-k-1)$. Model I has two calibrated parameters, whereas model II has four or five, depending on whether we prescribe or calibrate the initial peripheral temperature. 
For the individual warm-water trajectories, {\sf{AICc}} favors model II in all 12 swimmers, and the median {\sf{RMSE}} decreases from 0.273 to 0.119$^\circ$C. For the individual cold-water trajectories, {\sf{AICc}} favors model II in all four swimmers, and the median {\sf{RMSE}} decreases from 0.206 to 0.056$^\circ$C. To reduce the influence of temporal correlation in the digitized observations, we repeat the calibrations with subsampled trajectories. At 5-minute resolution, {\sf{AICc}} continues to favor model II in all 12 warm-water swimmers. At 10-minute resolution, {\sf{AICc}} favors model II in 9 of 12 warm-water swimmers and 3 of 4 cold-water swimmers. These results demonstrate the greater descriptive flexibility of model II, although temporal sampling influences the {\sf{AICc}} comparisons, and the results do not establish independent predictive validity.
\vspace*{0.2cm}
%%%%%%%%%%%%%%%%%%%%%%%%%%%%%%%%%%%%%%%%%%%%%%%%%%%%%%%%%%%%%%%%%%%%%%%%%
\paragraph*{Personalized thermal exposure maps.}
%%%%%%%%%%%%%%%%%%%%%%%%%%%%%%%%%%%%%%%%%%%%%%%%%%%%%%%%%%%%%%%%%%%%%%%%%
The critical-water-temperature law identifies the {\it{direction}} of
thermal drift, while model II also predicts the {\it{time}} required to
reach a thermal state. We combine these two concepts in {\it{personalized
thermal exposure maps}} that resolve water temperature and swim pace.
The maps report first-passage times to mild hypothermia at
$T_{\rm c}=\SI{36.0}{\degreeCelsius}$ and mild hyperthermia at
$T_{\rm c}=\SI{38.5}{\degreeCelsius}$ and thereby extend the thermal
phase boundary into an exposure-time landscape.
When calculating personalized thermal exposure maps, surface area follows the Du Bois relation in Eq.~(\ref{eq:S3}), environmental conductance follows $H=h_{\rm eff}A$, and core--peripheral conductance scales with area,
$G=G_{\rm ref}\,{A}/{A_{\rm ref}}$.
The implementation uses
$G_{\rm ref}=\SI{76.7}{W\,K^{-1}}$ and $\nu_{\rm c}=0.379$ as
calibration-informed reference values. We initialize the peripheral
temperature as
$T_{\rm p}^{0}=T_{\rm c}^{0}-{Q}/{G}$,
which enforces $\dot T_{\rm c}(0)=0$ and generates the initial
core-temperature plateau. We calculate first-passage times on a
30-second time grid and display exposure classes through 240~minutes.

First, we personalize the map to the {\it{marathon swimmer}} from
Figure \ref{fig05} (left) of the main text. This swimmer has a mass of
$\SI{95.5}{kg}$, a height of $\SI{185}{cm}$, and a BMI of
$27.9\,\si{kg/m^2}$. His total 37-km swim lasted 9~h~52~min, which corresponds
to an average pace near 1:36~min/100~m. We calibrate model II to his
measured cold-water trajectory and then vary pace while keeping his
body size fixed.

Figure~\ref{fig07} reveals a comparatively weak pace dependence across
much of the physiologically relevant range. The swimmer's large
mass-to-surface-area ratio limits environmental heat exchange relative
to his thermal mass and metabolic heat production. As a result, changes
in pace shift the thermal boundaries, but the exposure landscape
remains comparatively stable over a broad range of swimming speeds.
This behavior provides a direct example of personalization: for this
large, well-insulated swimmer, body geometry strongly moderates the
effect of swim intensity.

Next, we consider a {\it{recreational swimmer}} for comparison. 
The swimmer has a mass of $\SI{68}{kg}$, height of $\SI{175}{cm}$, and BMI of
$22.2\,\si{kg/m^2}$. 
We assign a reference pace of
1:50~min/100~m and calibrate the heat-production scale such that the
core reaches $\SI{36.0}{\degreeCelsius}$ after a 30~min non-wetsuit swim in
$\SI{18}{\degreeCelsius}$ water. 
Figure \ref{fig08} varies the swimmer's pace over the same range
as for the previous swimmer. 
The contrast between Figures ~\ref{fig07} and \ref{fig08} exposes the central role of {\it{personalization}}. Pace shifts thermal boundaries of the smaller recreational swimmer much more strongly than those of the larger marathon swimmer. His lower BMI and smaller mass-to-surface-area ratio increase environmental heat exchange relative to thermal mass, while changes in pace directly alter metabolic heat production. Faster swimming therefore moves the cooling--heating boundary toward colder water and substantially extends cold-water exposure times, whereas slower swimming shifts the same
swimmer toward earlier cooling. This simple example shows that the same water temperature and swim duration can consequently represent very different thermal exposures for different swimmers.

These maps illustrate the practical consequence of the model
hierarchy. The governing heat balance defines the phase boundary,
core--peripheral physiology determines the time scale, and swimmer
characteristics determine where that boundary lies. A fixed water
temperature therefore represents only one coordinate of a
swimmer-specific thermal exposure landscape.
\vspace*{0.2cm}
%%%%%%%%%%%%%%%%%%%%%%%%%%%%%%%%%%%%%%%%%%%%%%%%%%%%%%%%%%%%%%%%%%%%%%%%%
\paragraph*{Interpretation of model hierarchy.}
%%%%%%%%%%%%%%%%%%%%%%%%%%%%%%%%%%%%%%%%%%%%%%%%%%%%%%%%%%%%%%%%%%%%%%%%%
The three models answer distinct questions: Model~I identifies the governing heat balance and yields the critical-water-temperature scaling law analytically. Model II introduces an internal physiological degree of freedom and thereby resolves multiple thermal time scales. Model III allows geometry, swim intensity, and insulation to move the phase boundary between cooling and heating. The hierarchy deliberately preserves the simplest model whenever its assumptions suffice: additional physiological structure improves the description of non-monotonic trajectories, but it is not required to expose the fundamental scaling law.

\vspace*{0.2cm}
%%%%%%%%%%%%%%%%%%%%%%%%%%%%%%%%%%%%%%%%%%%%%%%%%%%%%%%%%%%%%%%%%%%%%%%%%
\paragraph*{Acknowledgments.}
%%%%%%%%%%%%%%%%%%%%%%%%%%%%%%%%%%%%%%%%%%%%%%%%%%%%%%%%%%%%%%%%%%%%%%%%%
The authors acknowledge
inspiration from the GTN Show 
and support from 
the Wu Tsai Human Performance Alliance,
the NSF CMMI grant 2320933, and 
the ERC Advanced Grant 101141626.

%%%%%%%%%%%%%%%%%%%%%%%%%%%%%%%%%%%%%%%%%%%%%%%%%%%%%%%%%%%%%%%%%%%%%%%%%


\begin{thebibliography}{99}
%%%%%%%%%%%%%%%%%%%%%%%%%%%%%%%%%%%%%%%%%%%%%%%%%%%%%%%%%%%%%%%%%%%%%%%%%

\bibitem{Williams2015}
C. L. Williams, G. L. Hagelin, G. L. Kooyman,
Hidden keys to survival: The type, density, pattern and functional role of emperor penguin body feathers,
%\emph{Proc. R. Soc. B.} 
Proc. R. Soc. B. \textbf{282}, 20152033 (2015).

\bibitem{TiptonBradford2014}
M.~Tipton, C.~Bradford,
Moving in extreme environments: Open water swimming in cold and warm water,
%\emph{Extreme Physiology \& Medicine} 
Extreme Phys. Med. \textbf{3}, 12 (2014).

\bibitem{Nadel1984}
E.~R.~Nadel,
Energy exchanges in water,
%\emph{Undersea Biomedical Research} 
Undersea Biomed. Res. \textbf{11}, 149 (1984).

\bibitem{Cannon1960}
P. Cannon, W. R. Keatinge,
The metabolic rate and heat loss of fat and thin men in heat balance
in cold and warm water,
J. Physiol. \textbf{154}, 329--344 (1960).

\bibitem{Nadel1974}
E.~R.~Nadel, I.~Holm\'er, U.~Bergh, P.-O.~{\AA}strand, and J.~A.~J.~Stolwijk,
Energy exchanges of swimming man,
%\emph{Journal of Applied Physiology} 
J. Appl. Phys. \textbf{36}, 465--471 (1974).

\bibitem{Markey2026}
K.~Markey, N.~Galan-Lopez, C.~Esh, S.~Carter, B.~Chrismas, M.~Mountjoy,
N.~Constantini, and L.~Taylor,
Thermoregulatory responses in open water and pool swimming: Presentation of hypothermia and hyperthermia within and outside of World Aquatics water temperature thresholds,
%\emph{Journal of Science and Medicine in Sport} 
J. Sci. Med. Sport \textbf{29}, 28--41 (2026).

\bibitem{Tipton2017}
M.~J.~Tipton, N.~Collier, H.~Massey, J.~Corbett, and M.~Harper,
Cold water immersion: Kill or cure?,
%\emph{Experimental Physiology} 
Exp. Phys. \textbf{102}, 1335--1355 (2017).

\bibitem{Chalmers2021}
S.~Chalmers, G.~Shaw, I.~Mujika, \emph{et al.},
Thermal strain during open-water swimming competition in warm water environments,
%\emph{Frontiers in Physiology} 
Front. Physiol. \textbf{12}, 785399 (2021).

\bibitem{Saycell2019}
J.~Saycell, M.~Lomax, H.~Massey, and M.~Tipton,
How cold is too cold? Establishing the minimum water temperature limits
for marathon swim racing,
Brit. J. Sports Med. \textbf{53}, 1078--1084 (2019).

\bibitem{WorldAquatics2026}
World Aquatics,
\textit{Competition Regulations},
%Part Three: Open Water Swimming Rules
(World Aquatics, 2026).

\bibitem{Fourier1822}
J. Fourier,
\textit{Th\'eorie Analytique de la Chaleur}
(Firmin Didot, Paris, 1822).

\bibitem{Aoki2001}
K. Aoki and D. Kusnezov,
Fermi-Pasta-Ulam $\beta$ model: Boundary jumps, Fourier's law, and scaling,
%\emph{Physical Review Letters} 
Phys. Rev. Lett. \textbf{86}, 4029--4032 (2001).

\bibitem{Skutnik2026}
B.~C.~Skutnik, M.~G.~Owen, M.~J.~Hite, L.~Sweitzer, J.~W.~Petersen,
B.~D.~Johnson, T.~D.~Mickleborough, J.~M.~Stager, and Z.~J.~Schlader,
Safe cold-water thresholds while wearing wetsuits approved for open
water swimming competitions,
%\emph{Journal of Applied Physiology} 
J. Appl. Physiol. \textbf{140}, 816--824 (2026).

\bibitem{Gagge1971}
A.P. Gagge, J.A.J. Stolwijk, and Y. Nishi,
An effective temperature scale based on a simple model of human physiological regulatory response.
Am. Soc. Heat. Refr. Air. Eng. \textbf{77}, 247--262 (1971).

\bibitem{Castellani2023}
M.P. Castellani, T.P. Rioux, J.W. Castellani, M.D. Reed, S. Whalen, M. Cisternelli, G. Python, M.J. Lewandowski, and X. Xu,
Validation of a human thermoregulatory model during prolonged immersion in warm water. 
Comp. Bio. Med. \textbf{167}, 107575 (2023).

\bibitem{Tarlochan2005}
F. Tarlochan S. and Ramesh, 
Heat transfer model for predicting survival time in cold water immersion, 
Biomed. Eng. Appl. Basis \& Comm. \textbf{17}, 159-–166 (2005).

\bibitem{Morton2025}
W.~J.~Morton, J.~Melau, R.~A.~Olsen, O.~M.~L{\o}vvik, J.~Hisdal,
and S.~S{\o}vik,
Thermal physiology of open water wetsuited swimming: A cohort study,
Temperature \textbf{12}, 245--263 (2025).

\bibitem{Markey2025}
K.~Markey, B.~C.~R.~Chrismas, C.~Esh, \emph{et al.},
A characterisation of elite open water swimmers' core temperature
responses and preparation practices in European Aquatics competitions,
%\emph{Journal of Science and Medicine in Sport} 
J. Sci. Med. Sport, in press,
doi:10.1016/j.jsams.2026.03.016.

\bibitem{Roxburgh2026}
B.~H.~Roxburgh,
Prolonged stable hypothermia during a 10-hour cold open-water marathon swim,
%\emph{Experimental Physiology} (2026),
Exp. Physiol. \textbf{111} 1652-1656 (2026).

\bibitem{Melau2019}
J.~Melau, M.~Mathiassen, T.~Stensrud, M.~Tipton, and J.~Hisdal,
Core temperature in triathletes during swimming with wetsuit in
10~$^\circ$C cold water,
Sports \textbf{7}, 130 (2019).

\bibitem{Miller1974}
N.C. Miller and R.C. Seagrave,
A model of human thermoregulation during water immersion.
Comput. Biol. Med. \textbf{4}, 165--182 (1974).

\bibitem{Tikuisis1988}
P. Tikuisis, R.R. Gonzalez, and K.B. Pandolf,
Thermoregulatory model for immersion of humans in cold water.
J. Appl. Physiol. \textbf{64}, 719--727 (1988).

\bibitem{Xu2005}
X. Xu, P. Tikuisis, R. Gonzalez, and G. Giesbrecht,
Thermoregulatory model for prediction of long-term cold exposure.
Comput. Biol. Med. \textbf{35}, 287--298 (2005).

\bibitem{Castellani2021}
M.P. Castellani, T.P. Rioux, J.W. Castellani, A.W. Potter, and X. Xu,
A geometrically accurate 3-dimensional model of human thermoregulation for transient cold and hot environments.
Comput. Biol. Med. \textbf{138}, 104892 (2021).

\bibitem{Xu2021}
X. Xu, T.P. Rioux, J. Gonzalez, E.O. Hansen, J.W. Castellani, W.R. Santee, et al.,
A digital tool for prevention and management of cold weather injuries---Cold Weather Ensemble Decision Aid (CoWEDA).
Int. J. Biometeorol. \textbf{65}, 1415--1426 (2021).

\end{thebibliography}
\end{document}